# Switchable quantized conductance in topological insulators revealed by the Shockley-Ramo theorem

Paul Seifert,[1,2] Marinus Kundinger,[1,2] Gang Shi,[3] Xiaoyue He,[3] Kehui Wu,[3] Yongqing Li,[3] Alexander Holleitner,[*,1,2] and Christoph Kastl[4]

[1]*Walter Schottky Institut and Physics department, Technical University of Munich, Am Coulombwall 4a, 85748 Garching, Germany.*

[2]*Nanosystems Initiative Munich (NIM), Schellingstr. 4, 80799 München, Germany.*

[3]*Institute of Physics, Chinese Academy of Sciences, Beijing 100190, China.*

[4]*Molecular Foundry, Lawrence Berkeley National Laboratory, 1 Cyclotron Road, 94720 Berkeley, California, United States.*

**\*holleitner@wsi.tum.de**

## Abstract

**Crystals with symmetry-protected topological order, such as topological insulators, promise coherent spin and charge transport phenomena even in the presence of disorder at room temperature.[1–4] Still, a major obstacle towards an application of topological surface states in integrated circuits is a clear, reliable, and straightforward read-out independent of a prevailing charge carrier density in the bulk. Here, we demonstrate how to image and read-out the local conductance of helical surface modes in the prototypical topological insulators $Bi_2Se_3$ and $BiSbTe_3$. We apply the so-called Shockley-Ramo theorem to design an optoelectronic probe circuit for the gapless surface states, and surprisingly find a precise conductance quantization at $1 \cdot e^2/h$.[5] The unprecedented response is a clear signature of local spin-polarized transport, and it can be switched on and off via an electrostatic field effect. The macroscopic, global read-out scheme is based on the displacement current resistivity, and it does not require coherent transport between electrodes.[6] It provides a generalizable platform for studying further non-trivial gapless systems such as Weyl-semimetals and quantum spin-Hall insulators.[7,8]**

# Main

In radiation detectors, signal formation often relies on the so-called Shockley-Ramo theorem,[9,10] which is distinct to the Landauer-Büttiker formalism typically used to describe mesoscopic charge transport from one electrode to another one.[11,12] Instead, radiation entering the detector locally creates free charge carriers in an overall insulating detector medium. The injected charges never reach an electrode, but by their local motion, a macroscopic displacement is electrostatically induced between the detector's electrodes. The macroscopic displacement signal is largely independent of the excitation position within the detector volume.[9,10] However, only those local current components contribute to the macroscopic signal, which align parallel to the so-called weighting field. The latter describes the overall electrostatic potential distribution for a specific detector geometry and electrode configuration.

Recently, Song *et al.* suggested that also for two-dimensional gapless metals, *i.e.* a conductive detector medium,[5] a locally excited photocurrent $j_{loc}(x,y)$ induces a macroscopic displacement signal with magnitude $I = A \int j_{loc}(x,y) \cdot \nabla \phi(x,y) \, dx \, dy$ (1), with the weighting field $\nabla \phi(x,y)$ derived from the potential distribution $\phi(x,y)$ within the investigated device, and $A$ considering the resistance of the overall circuitry. The weighting field coincides with the electrostatic field within a multi-terminal device in the absence of a transversal Hall conductivity.[5] While the Shockley-Ramo response is trivial for biased two-terminal devices or for materials with local anisotropies like potential fluctuations,[13] defects, or p-n junctions,[14] we reveal that the scheme also allows detecting local currents which are much more intrinsic in nature. In particular, we determine the local conductance of topological surface states in field effect devices with the help of a focused laser beam. Such a read-out scheme complements more conventional transport experiments which achieve the differentiation between surface and bulk conduction in topological insulators by either suppressing bulk conduction via electrostatic doping and the growth of materials with a reduced bulk conductivity,[15–17] or by selectively addressing the spin-helical structure of the topological surface states via optoelectronic methods.[18–20]

Figure 1a sketches our Shockley-Ramo detection scheme, which is based on an in-plane symmetry breaking in prototypical $Bi_2Se_3$-circuits on a $SrTiO_3$ substrate.[16,17] The $Bi_2Se_3$-film is contacted by two metallic source and drain contacts on the left and right side, but the weighting field $\nabla\phi$ is dominated by a gate potential $V_{gate} > 0$, applied to the back side of the $SrTiO_3$ substrate. Then, at the edges of the $Bi_2Se_3$-film, the weighting field distribution runs perpendicular to the edge, which breaks the in-plane symmetry of an otherwise isotropic local current $j_{loc}(x,y)$. In particular, when charges are locally added to the electron system, for example by optical excitation (red cone in Fig. 1a), the weighting field at the edges senses a net current flowing into the sample (gray arrow within the cone). There are simply no states flowing out of the $Bi_2Se_3$-film. Assuming this symmetry breaking, and the Fermi-energy to be within the Dirac cone of the topological insulator, one expects to measure the local quantized conductance of the surface state propagating into the sample, *i.e.* $1 \cdot e^2/h$. Since the materials are gapless, the added charges always end-up at the Fermi-energy. This argument also holds for a local interband photoexcitation after thermalization and relaxation of hot charge carriers. Importantly, the symmetry breaking is not achieved for ungated devices (Fig. 1b). There, the weighting field lines only extend from source to drain within the $Bi_2Se_3$-film. Consequently, this standard two-terminal geometry detects charges moving parallel to the edge. The corresponding global detector response averages out to zero, because locally there will be two states occupied with opposite directions (two gray arrows in red cone).

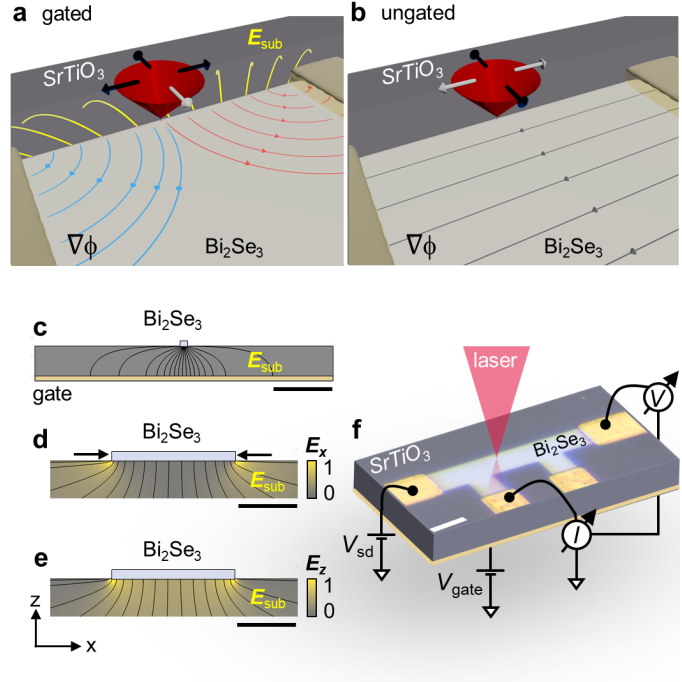

**Figure 1. Gated and ungated field-effect devices made from a topological insulator. a,** A field effect device is fabricated from a thin $Bi_2Se_3$-film on a $SrTiO_3$ substrate. A gate voltage $V_{gate}$ applied to the backside of the substrate controls the weighting field distribution $\nabla\phi$ aligned perpendicular to the edge of the $Bi_2Se_3$-film. A local excitation (red cone) generates a net current perpendicular to the edge (gray arrow) coupling to the source-drain electrodes (left and right). **b,** Without the gate, only the currents parallel to the edge (gray arrows in red cone) are coupled to the electrodes through $\nabla\phi$. **c,** Side view of simulated electric field between the millimeter-sized global backgate and the $Bi_2Se_3$-microstructure. Scale bar, 1 mm. **d,** and **e,** Magnified view of the anisotropic electric field $\boldsymbol{E}_{sub}$. The in-plane ($E_x$) and out-of-plane ($E_z$) fields are in units of $10^7$ V/m for $V_{gate} = 100$ V. Black lines indicate the electric field. Scale bars, 5 μm. **f,** π-shaped three-terminal device with external circuitry. Scale bar, 5 μm.

The anisotropic field distributions as in Fig. 1a can be realized, when we gate thin films of $Bi_2Se_3$ and $BiSbTe_3$ with a macroscopic backgate by the help of the thick, high-$k$ $SrTiO_3$ substrate. Since the lateral footprint of the films is much smaller than the extension of the bottom gate (Fig. 1c), this particular device geometry does not resemble a parallel plate capacitor. The device is rather reminiscent of a plate-wire configuration featuring a highly anisotropic field distribution of the electric field $E_{sub}$ within the substrate. Importantly, the dielectric constant abruptly changes by orders of magnitude at the vacuum/SrTiO$_3$ interface, which aligns the field parallel to the interface near the edges of the Bi$_2$Se$_3$-films. A close-up of the field distribution demonstrates a corresponding enhancement of the in-plane field component $E_x$ near the edges of the films (arrows in Fig. 1d). We note that the simulated peak field strength is in the order of $10^7$ Vm$^{-1}$ at $V_{gate}$ = 100 V with an assumed dielectric constant of $\varepsilon_{SrTiO3} = 10^4$ at 5 K.[21] This field strength certainly affects the charge carrier motion in the investigated topological films. However, it is orders of magnitude smaller than the fields required to modify the band structure on an atomistic level. Figure 1e shows the out-of-plane field component $E_z$. As expected for a bottom gate geometry, it extends below the footprint of the Bi$_2$Se$_3$-film with maxima at the edges. Figure 1f depicts an optical image of a π-shaped four-terminal device made from a Bi$_2$Se$_3$-film and the overall electronic circuitry. We locally inject charges at the edge of the Bi$_2$Se$_3$-film using a focused laser at energy $E_{photon}$ = 1.5 eV (red cone). Two of the electrodes are contacted to low-impedance contacts to measure the macroscopic current signal $I$, and they act as a reference potential for the backgate electrode. A high impedance differential voltage amplifier is wired to a third contact for the simultaneous measurement of the macroscopic voltage $V$. When we scan the laser position across the device, the three-terminal circuitry allows us to determine a local conductance $G(x,y) = I(x,y)/V(x,y)$ for the photogenerated carriers for each position $x$ and $y$.

Figures 2a and 2b show the spatially resolved current $I(x,y)$ and voltage $V(x,y)$ of an *n*-type Bi$_2$Se$_3$-device for $V_{gate}$ > 0 V at zero bias and $T$ = 4.2 K. The current is measured between the contacts labelled $S$ and $D$ (Fig. 2a), while the voltage is concurrently measured between the contacts labelled $V^+$ and $V^-$ (Fig. 2b and methods). Within our spatial resolution given by the laser spot size ~1-2 μm, we detect a distinct conductance at the edges of the device (dashed lines). Fig. 2c depicts the resulting conductance map $G(x,y)$ of the π-shaped device. Averaging across all positions on the circuit, we find a conductance peak with mean value $|G| = 0.94 \cdot e^2/h$ and full width half maximum $\Delta G = 0.24 \cdot e^2/h$ (dashed orange distribution in Fig. 2d) on top of a broad background. The peak value of the orange distribution near $1 \cdot e^2/h$ implies that the corresponding local transport is carried by a spin-polarized mode of the topological surface state. Figures 2e-h present corresponding data and conductance histogram for $V_{gate} < 0$ V (c.f. Supplementary Figure 1). We observe that the edge response is suppressed for this gate setting, as one would naively expect for *n*-type films (Supplementary Figure 2). In a previous study,[13] we demonstrated that then, a photo-thermoelectric current dominates, driven by local fluctuations of the Seebeck coefficient in the surface states. In the present experiment, this effect can slightly be resolved within the noise level (Fig. 2e - 2g). The conductance map $G(x,y)$ appears to be random (Fig. 2g), and the overall conductance histogram is distributed around $|G| = 0$ (Fig. 2h), which also explains the background distribution in Fig. 2d. The sign of $G(x,y)$ is determined by the local current direction.

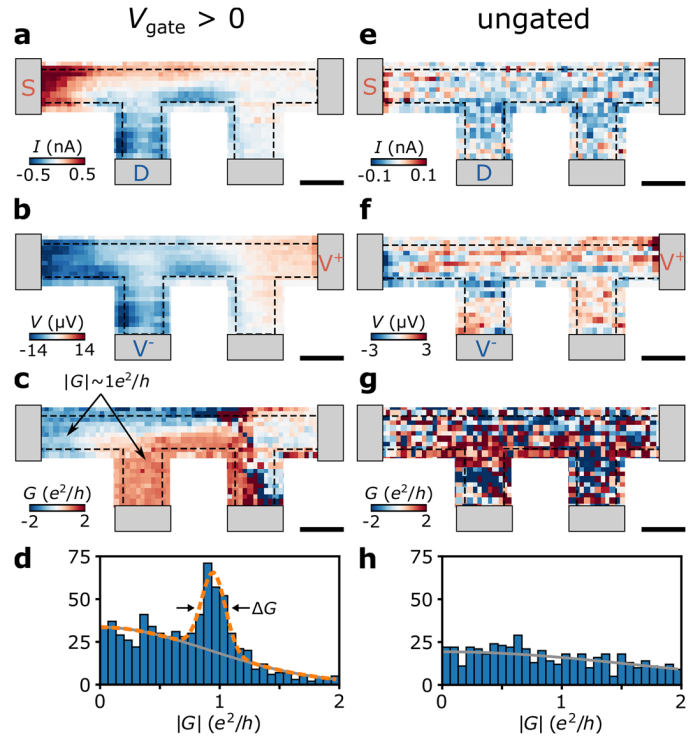

**Figure 2. Switchable quantized conductance for *n*-type Bi$_2$Se$_3$-circuits. a-c,** Spatial maps of the current *I*, the voltage *V*, and the local conductance $G = I/V$ at a positive gate voltage (data for $V_{gate} = 15$ V). The laser-induced current (voltage) is measured between the contacts labeled *S* and *D* ($V^+$ and $V^-$). **d,** The histogram across all positions shows a defined conductance peak (orange dashed line) with mean value $|G| = 0.94 \cdot e^2/h$, a full width half maximum $\Delta G = 0.24 \cdot e^2/h$, and a broad background signal centered around zero conductance (gray line). **e-g,** For negative and zero gating, the quantized conductance detection is switched

off (data for $V_{gate}$ = -30 V), and the response is dominated by potential fluctuations. **h,** The conductance is (normally) distributed around $|G|$ = 0 (gray line). The dashed lines indicate the edge of the circuit. Experimental parameters are $E_{photon}$ = 1.5 eV and $T_{bath}$ = 4.2 K. Scale bars, 5 µm.

According to the Shockley-Ramo theorem, different configurations of floating and grounded electrodes will change the weighting field and accordingly the macroscopic displacement response.[5,9,10] Figure 3 depicts the simulated weighting fields (black lines in Figs. 3a-c) and the corresponding measured current (Fig. 3d-f) for different circuit configurations of the π-shaped four-terminal device. In all cases, source (*S*) and drain (*D*) are grounded, and no bias is applied. All other contacts are floating. For the simulations,[5] we apply boundary conditions such that the weighting field lines terminate perpendicularly at both the sample edges and the contact electrodes.[5] The simulations do not consider a self-consistent treatment of the screening within the metallic surface states. However, the qualitative picture remains the same, since the main effect of such a screening would be to decrease the characteristic length scale over which the field component perpendicular to the edge decays. The red and blue arrows in Figs. 3a indicate locally excited currents $j_{loc}(x,y)$ resulting from the in-plane symmetry breaking at the edges and flowing perpendicular to the boundaries of the π-shaped device. These local currents couple via the weighting field either to the source electrode (red colored arrows) or to the drain electrode (blue colored arrows). According to equation (1), this coupling determines the sign of the globally sensed signal. A comparison of the simulations and experiments clearly demonstrates that the measured signal is consistent with the expected Shockley-Ramo response. In particular, in Figs. 3d and 3e, we can accurately explain the non-local behavior of the negative current (blue) between contact *D* and the floating contact next to it. It also explains how local currents $j_{loc}(x,y)$ with opposite polarity (indicated by direction of arrows) at opposite edges results in the same polarity of the global current *I(x,y)*. Experimentally, there is a certain degree of asymmetry in the contacts with respect to the magnitude of the current signal. This is apparent for *S* and *D* in Fig. 3f, and it is most likely caused by the varying contact resistances we find for the different electrodes (data not shown).

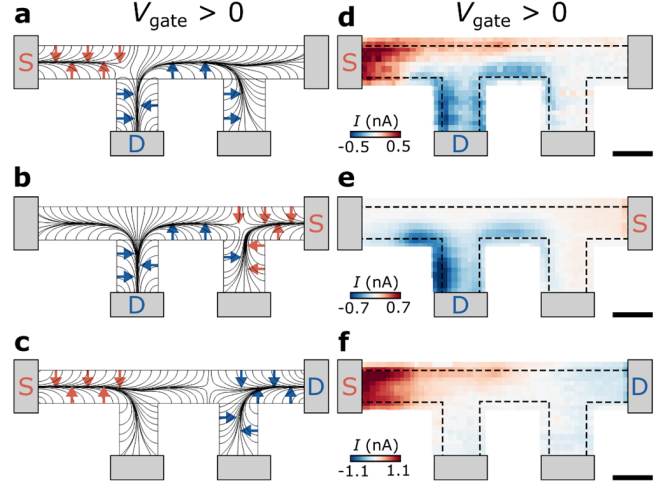

**Figure 3. Shockley-Ramo circuit configurations. a,-c,** Simulated field distribution and **d,-f,** measured current signal on a $Bi_2Se_3$-circuit for different electrode configurations. *S* and *D* denote the contacts wired as source and drain, respectively. The other contacts are floating. The arrows indicate the direction of the local current $j_{loc}(x,y)$. The sign of the globally measured signal is determined by the coupling of $j_{loc}(x,y)$ into *S* and *D* mediated by the weighting field distribution between the contacts. Red (blue) arrows indicate a global current into *S* (*D*). The dashed lines indicate the edge of the circuit. Experimental parameters are $E_{photon}$ = 1.5 eV and $T_{bath}$ = 4.2 K. Scale bars, 5 µm.

For describing the quantized conductance in terms of a Landauer-Büttiker formalism, one would measure $I(x,y)$ with the help of two contacts and utilize the two remaining contacts of the π-shaped four-terminal circuit to probe the corresponding voltage $V(x,y)$.[6] Surprisingly, for all such standard four-terminal wirings, we could never concurrently detect a finite *I* and *V* for any position *x* and *y* in the circuits to determine $G(x,y)$ (data not shown). Therefore, in our present understanding, we cannot apply the Landauer-Büttiker formalism. Instead, Fig. 3c explains why we cannot measure a signal in a standard four-terminal wiring: according to the Shockley-Ramo weighting field, there is no position $(x,y)$ in the circuit which connects to all four probes (the same applies to all further wiring possibilities as in Figs. 3a and 3b). Actually, we exploit this insight in the three-terminal wiring as in Fig. 1f to concurrently resolve $I(x,y)$ and $V(x,y)$ at any given position.

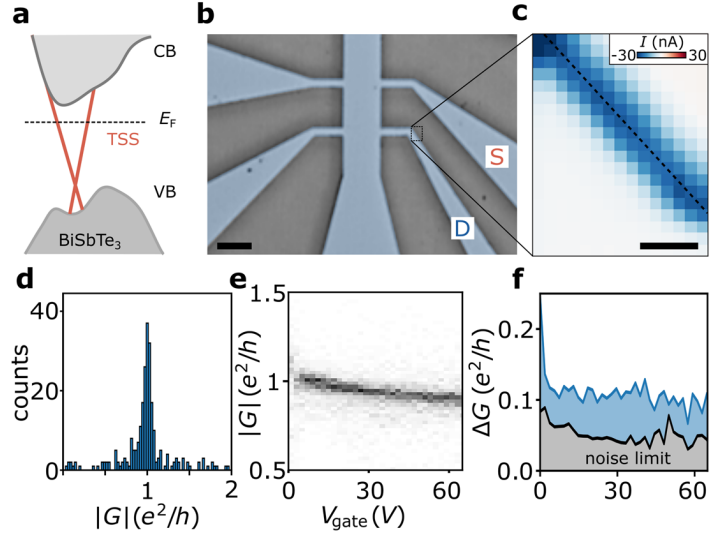

**Figure 4. Enhanced quantization in bulk insulating BiSbTe$_3$. a,** Schematic band structure of BiSbTe$_3$ with conduction band (CB), valence band (VB), topological surface state (TSS), and Fermi-level $E_F$. **b,** Macroscopic BiSbTe$_3$ Hall bar with source (S) and drain (D) electrodes. Scale bar, 50 µm. **c,** Laser-induced current map of the area indicated by the rectangle in **b** ($V_{gate}$ = 7.4 V). The current is localized at the sample edge (dashed line). Scale bar, 1 µm. **d,** Conductance histogram for the measurement depicted in **c** with $|G|$ = 1.003·$e^2/h$ and $\Delta G$ = 0.1·$e^2/h$. **e,** $|G|$ as function of applied gate voltage $V_{gate}$. The dotted line marks 1·$e^2/h$. **f,** Distribution of the conductance width $\Delta G$ (blue curve) and noise limit (black curve). Experimental parameters are $E_{photon}$ = 1.5 eV and $T_{bath}$ = 4.2 K.

The so-far discussed, as-grown Bi$_2$Se$_3$-films are typically *n*-doped due to defects, and the Fermi-level is situated above the bulk gap for positive gate voltages.[17,22] For such *n*-doped samples, hexagonal warping of the surface state dispersion and the effective coexistence of surface states with bulk states open additional scattering channels.[23–25] In this way, we explain the reduced mean value $|G| < 1·e^2/h$, the rather broad $\Delta G$, and the broad background signal as observed in Fig. 2d. Therefore, we also study BiSbTe$_3$-films, where the Fermi-level is in the bulk gap for $V_{gate}$ = 0 V, albeit not necessarily at the Dirac point (Fig. 4a).[26] Furthermore, to better differentiate the edge positions from the interior parts of the circuits, we macroscopic Hall-bar structures as depicted in Fig 4b. Again, also for such devices, the signal generation is clearly localized at the sample edges (Fig. 4c and Supplementary Fig. 3). The corresponding histogram of $|G|$ shows a sharp quantization with mean value $|G| = 1.003 \cdot e^2/h$ and much reduced width of $\Delta G = 0.1 \cdot e^2/h$ (Fig. 4d). The quantized conductance appears at $V_{gate} > 0$ V at $|G| = 1 \cdot e^2/h$ (Fig. 4e), and it gradually decreases to $|G| \approx 0.9 \cdot e^2/h$ for an increased $n$-type gating. Again, we interpret this finding to be a signature of increased scattering or hybridization between bulk and surface states. We note that for $V_{gate} > 50$ V, the gate capacitance decreases, especially under illumination, which explains the saturating behavior of the conductance at large gate voltages. To quantify the sharpness of the quantization, we plot $\Delta G$ together with the calculated noise limit for each measurement (Fig. 4f). The experimental noise limit is determined by measuring the rms-noise of the current $\Delta I$ and voltage $\Delta V$ and by calculating the expected Gaussian broadening of the conductance distribution as $\Delta G^2 = (\Delta I/V)^2 + (\Delta V \cdot I/V^2)^2$ for each individual measurement. From the difference between the measured and the calculated curves, we can estimate that the conductance is defined better than $\Delta G = 0.05 \cdot e^2/h$ for all gate voltages. These results establish that the conductance coincides with the value $1 \cdot e^2/h$ within the experimental resolution for an extended range of gate voltages. Our experiments suggest that within the Shockley-Ramo detection scheme, a coherent charge and spin transport between the excitation spot and the contact is not a prerequisite for detecting a local quantized conductance. In particular, the lateral spacing of our circuits exceeds by far the surface states' coherence length (~100 nm).[27] Yet, it is crucial to utilize a focused laser spot. In particular, we do not detect a finite conductance signal for a defocused global excitation of the overall circuits (data not shown). In our understanding, the weighting field at the edges acts as a directional momentum filter perpendicular to the edges independent of the microscopic processes during photoexcitation and relaxation of the hot charge carriers.[28] A focused laser spot then reveals this symmetry breaking just at one particular edge and at one certain position, while a defocused laser excites charge carriers at many positions and edges averaging out the macroscopic response. We can further exclude a photogalvanic effect and a spin-Hall photoconductance as the underlying conductance mechanisms, since the measured signal is independent of the polarization of the laser excitation (Supplementary Figure 4).[18,19,29] The quantized conductance is observable up to $T \sim 14$

K, and we find that in this range, the quantized value is independent of temperature (Supplementary Figure 5), which suggests that the microscopic mechanism is different from the recently predicted 'squeezed edge currents' in multi-valley insulators.[30]

The optical excitation occurs at a photon energy of ~1.5 eV via interband transitions across the bulk gap involving surface and bulk states.[31] In our point of view, the observed transport at $e^2/h$ occurs predominantly at the Fermi-level within the laser spot, because the thermalization and the overall relaxation processes of photoexcited charge carriers occur on a sub-picosecond timescale.[32] The local conductance response is governed by the interplay of different length scales, which are the Thomas-Fermi-screening length (few nanometers), the inelastic mean free path (~10-100 nm), the diffusion length of hot charge carriers (several 100s nm), and the already mentioned laser spot size (1-2 µm). In our understanding, the excited surface state population locally increases the chemical potential according to the compressibility of the surface states, and it can persist up to several hundreds of picosecond.[33,34] Particularly, for the $BiSbTe_3$-films, the Fermi-level is within the surface states, such that the quantized conductance of the surface states dominates. This transport is then sensed macroscopically through the weighting field $\vec{\nabla}\phi(x,y)$ at the source and drain contacts of the devices (Fig. 1). Ultimately, we expect the smallest relevant length scale to be the screening length of the hard-wall potential at the sample edges.

Within the Shockley-Ramo framework, our model can successfully explain the switching behavior of the photoresponse, which is at first sight counterintuitive due to the gapless nature of the surface states. Furthermore, we can accurately predict the polarity, the long-range character, and the apparent non-locality of the conductance. To gain further insights into the microscopic origins, it will be necessary to disentangle the effects of electric field enhancement at the edges, the Thomas-Fermi screening, the potential fluctuations, and the gating of bottom vs. top surface state. In our point of view, circuits with additional top-gates made from graphene with an hBN spacer may help to differentiate between bottom and top surface as well as to tune the films into a quantized conductance regime also for $V_{gate} < 0$. Furthermore, scanning optical near field-measurements at lower photon energies allow for further exploring the underlying optoelectronic processes at the relevant lateral length scales,[35] but this nanoscopic investigation is beyond the current manuscript. Currently, the 'lateral resolution' of the quantized conductance imaging is limited to about 300 nm (Supplementary Figure 6).

Overall, we demonstrate a novel optoelectronic detection scheme that applies the Shockley-Ramo theorem to electric conductors, which allows us to microscopically probe, yet macroscopically read-out the intrinsic quantized conductance of topological surface states. We anticipate that this read-out scheme provides a generalizable platform for studying local transport phenomena in further non-trivial gapless systems such as graphene, Weyl-semimetals or quantum spin-Hall insulators.[7,8,35]

## Methods

*Growth and lithography parameters of $Bi_2Se_3$ and $BiSbTe_3$-films.* The $Bi_2Se_3$ ($BiSbTe_3$) thin films are grown in a home-made molecular beam epitaxy (MBE) system with a base pressure better than $1\times10^{-10}$ mbar. Prior to the growth, the $SrTiO_3$(111) substrates are boiled in deionized water at 85 °C for 50 minutes, then annealed in pure $O_2$ environment at 1050 °C for 2 hours in order to obtain smooth surfaces. During the growth, high purity Bi (99.997%) and Se (99.999%) for $Bi_2Se_3$ and Bi (99.999%), Sb (99.999%) and Te (99.999%) for $BiSbTe_3$ are evaporated from Knudsen cells. A quartz crystal thickness monitor is used to calibrate the flux rate. A Bi/Se (Bi/Te) flux ratio of about 1:10 is kept to ensure Se-rich (Te-rich) growth condition. The Bi:Sb composition ratio is varied by adjusting Sb flux. The deposition rate is about 0.25 nm/min. The substrate temperature is maintained at 360°C throughout film growth. The $Bi_2Se_3$ (10 nm film thickness) samples are in-situ capped by 30 nm Se. The $BiSbTe_3$ film thickness is 15 nm. The films are lithographically patterned into 50 μm wide Hall bars or into 5 μm wide π-shaped bars using reactive ion etching with an Ar flow rate of 40 sccm and a coil power of 80 W. The etching rate is about 1 nm/s. Cr/Au (3 nm/30 nm) layers are then deposited on the back of the $SrTiO_3$ substrates and the top surfaces of the films to serve as backgate electrodes and ohmic contacts to the $Bi_2Se_3$ ($BiSbTe_3$) thin films, respectively. The Se-capping layer is removed by heating the samples to 120°C for 20 min. The quantized conductance was reproduced on four different samples from different growth processes.

*Low temperature scanning conductance microscopy.* Laser-induced current and voltage experiments were performed using a confocal laser scanning microscope with a diffraction limited spatial resolution FWHM ≈ 1.3 μm. We use pulsed excitation with ≈100 ps pulses at a 40 MHz repetition rate and a photon energy 1.5 eV. The optical excitation energy is larger than the band gap of $Bi_2Se_3$ and $BiSbTe_3$, but much smaller than the band gap of the $SrTiO_3$ substrate. The results are reproduced with *cw*-excitation. The current *I* at each excitation position is measured via a low-impedance transconductance amplifier (*DL instruments 1211*, typical gain $10^7$) at the drain contact. The transconductance amplifier provides a virtual ground potential at drain. At the same time, a voltage source (*Yokogawa 7651*) maintains zero bias between source and drain contact. The voltage *V* is read out via a high impedance differential amplifier (*Femto DLPVA-100-F-D*, typical gain $10^3$, 1 TΩ input impedance), which draws negligible input current. The sign of the voltage is given according to $\phi(V+)-\phi(V-)$. A global backgate $V_{gate}$ adjusts the electric field. All other contacts are floating. We note that, for an excitation near the source (drain) contact (*S* and *D* in Fig. 2a), a positive current indicates a predominant coupling into *S* (*D*). Importantly, the magnitude of the voltage (Fig. 2b) is likewise consistent with the geometric coupling to the source and drain contacts: an electron current into *D* increases the potential of $V^-$ and hence results in a negative voltage. Similarly, an electron current into *S* increases the potential of S, which is detected again as a negative voltage, since *S* and drain $V^-$ are actively maintained at the same potential (cf. Fig. 1f). We use DC current/-voltage measurements, since kHz modulated lock-in measurements suffer from spurious background signals. The experiments are conducted in a bath cryostat at temperatures $T$ = 4.2 K – 14 K in a He atmosphere at 10 mbar pressure. The optical power was varied between 1 µW - 100 µW.

*Electric and weighting field simulations.* All simulation results are obtained using the electrostatics module of the commercial finite-element simulation software COMSOL Multiphysics, version 4.4. For the anisotropic backgate field distribution, the capacitor structure is modelled with a dielectric constant of $10^4$ for the ferroelectric SrTiO3 at T ≈ 5 K and an applied bias voltage of 100 V between the backgate electrode and a 10 µm microstructure. The weighting field lines in the π-shaped bars are modeled according to Song et al. as the potential gradient in the microstructure in absence of a Hall conductivity. An elevated fixed potential at the sample boundaries owing to the anisotropic backgate field distribution was assumed as boundary condition. The source and drain contacts are simulated as fixed zero potential and ground respectively, while the other pair of electrodes are simulated as floating potential. All modelling results are obtained from a two-dimensional simulation geometry and a stationary solver.

## Data availability

The data that support the findings of this study are available from the corresponding author upon reasonable request.

**Acknowledgements**

We thank J. Song for in-depth discussions. This work was supported by the DFG via SPP 1666 (grant HO 3324/8), ERC Grant NanoREAL (n°306754), the Center of NanoScience (CeNS) in Munich, and the Munich Quantum Center (MQC). C. K. acknowledges funding by the Molecular Foundry supported by the

**Corresponding author**

holleitner@wsi.tum.de

# Supplementary Information

# Switchable quantized conductance in topological insulators revealed by the Shockley-Ramo theorem

Paul Seifert,[1,2] Marinus Kundinger,[1,2] Gang Shi,[3] Xiaoyue He,[3] Kehui Wu,[3] Yongqing Li,[3] Alexander Holleitner[*1,2], Christoph Kastl[4]

[1]*Walter Schottky Institut and Physics department, Technical University of Munich, Am Coulombwall 4a, 85748 Garching, Germany.*
[2]*Nanosystems Initiative Munich (NIM), Schellingstr. 4, 80799 München, Germany.*
[3]*Institute of Physics, Chinese Academy of Sciences, Beijing 100190, China.*
[4]*Molecular Foundry, Lawrence Berkeley National Laboratory, 1 Cyclotron Road, 94720 Berkeley, California, United States.*
**[*]holleitner@wsi.tum.de**

# Supplementary Figure 1

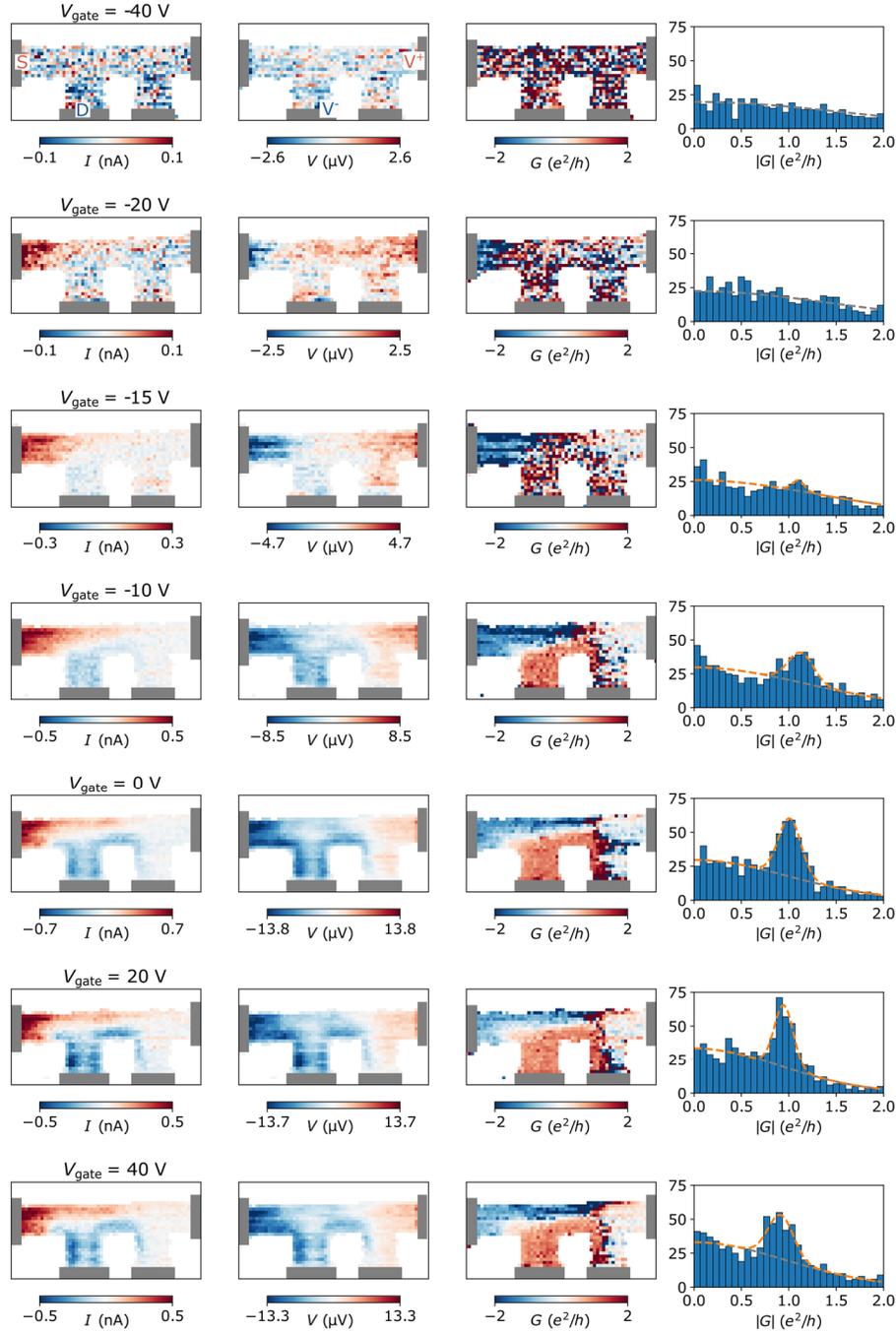
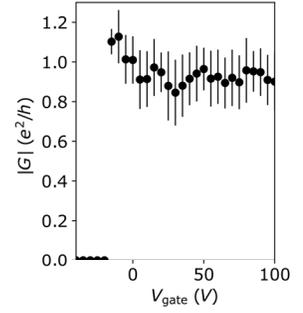

**Supplementary Figure 1. Switching behavior of quantized conductance in *n*-type Bi$_2$Se$_3$ (π-bar device). (a)** From left to right: maps of the current $I_{photo}$, voltage $V_{photo}$ and local conductance $G_{photo}$ and conductance histogram for gate voltages from $V_g$ = -40 V to $V_g$ = 40 V. The current was measured between the electrodes S and D and the voltage between the electrodes V+ and V- (as indicated in the topmost panels). **(b)** Mean value of the conductance distribution associated with the edge current. Error bars indicate the Gaussian width of the distribution. Experimental parameters are $E_{photon}$ = 1.5 eV, $T$ = 4.2 K.

## Supplementary Figure 2

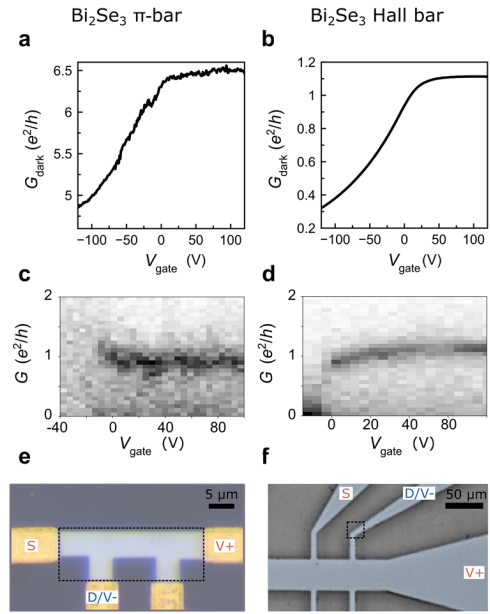

**Supplementary Figure 2. Dark conductance $G_{dark}$ and local conductance $G$ of $Bi_2Se_3$ films.** Dark conductance measured in standard two-terminal geometry of **a,** 10 nm $Bi_2Se_3$, π-shaped micro bar device. **b,** 10 nm $Bi_2Se_3$, macroscopic Hall bar device. Local conductance measured in Shockley-Ramo three-terminal geometry of **c,** 10 nm $Bi_2Se_3$, π-shaped micro bar device.d**e,** 10 nm $Bi_2Se_3$, macroscopic Hall bar device. **e, f,** Optical images of the corresponding device structures. The dark conductance was measured between S and D using a standard two-terminal resistance measurement. The local conductance was measured in a three-terminal configuration as outlined in the main manuscript: The current was measured between the unbiased electrodes labelled S and D. Voltage was measured concurrently between the electrodes labelled V+ and V-. The dashed rectangles mark the image area from which the local conductance values were determined. Experimental parameters are $E_{photon}$ = 1.5 eV, $T$ = 4.2 K.

## Supplementary Figure 3

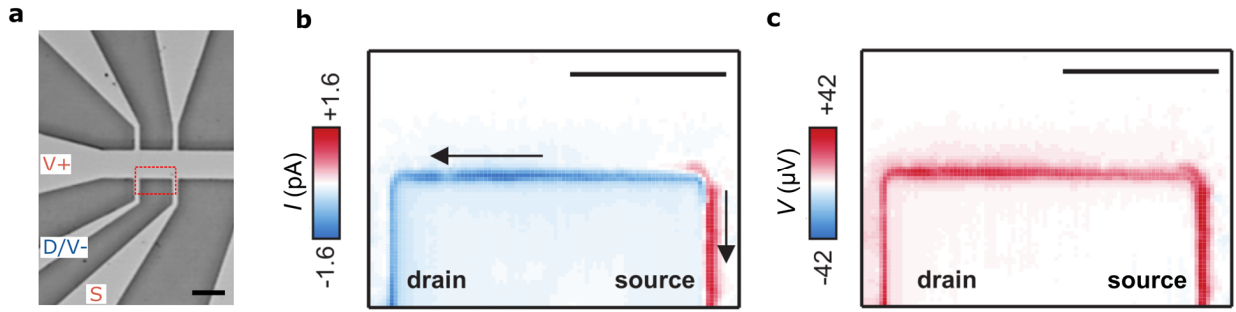

**Supplementary Figure 3. Shockley-Ramo photocurrent in macroscopic $BiSbTe_3$-Hall bar circuits. a,** Optical microscope image of 15 nm thin $BiSbTe_3$-film (gray) on $SrTiO_3$ (dark gray). Scale bar is 50 µm. The current was measured between the source (S) and drain (D) contact. The voltage was measured between the electrodes labelled V+ and V-. Simultaneously recorded current- **(b)** and voltage-maps **(c)** ($V_{gate}$ = 40 V). Positive (negative) current $I_{photo}$ corresponds to a local current coupling to source (drain) as indicated by the arrows. Experimental parameters are $E_{photon}$ = 1.5 eV, $T$ = 4.2 K.

**Supplementary Figure 4**

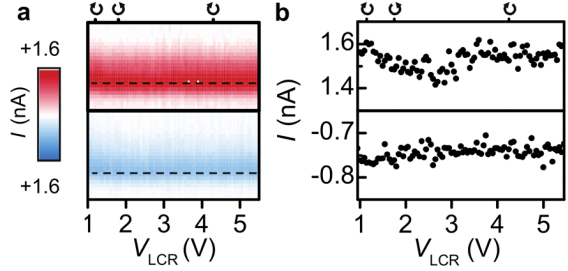

**Supplementary Figure 4. Polarization dependence of Shockley-Ramo respsonse in BiSbTe$_3$ (Hall bar device). a,** Photocurrent perpendicular to circuit edge (dashed line) for varying photon polarization. Top panel and lower panel show different edges with positive and negative photocurrent. A nematic liquid crystal retarder (LCR) adjusts the polarization. The retardance and hence the polarization depends non-linearly on the voltage $V_{LCR}$ applied to the liquid crystal. The top symbols denote circularly right-handed and circularly left-handed polarized excitation. **b,** The photoresponse is independent of the polarization within the noise level. Experimental parameters are $E_{photon}$ = 1.5 eV, $T$ = 4.2 K.

**Supplementary Figure 5**

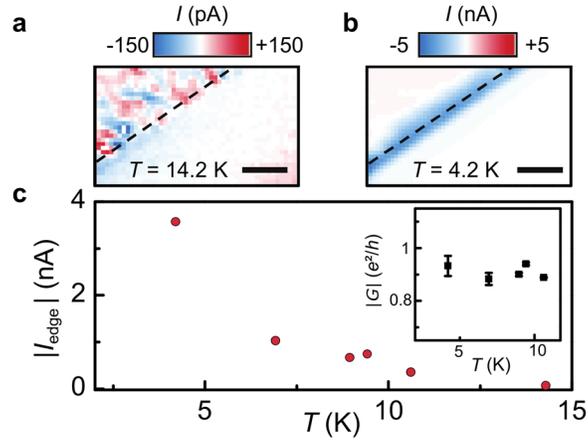

**Supplementary Figure 5. Temperature dependence of Shockley-Ramo response in BiSbTe$_3$ (Hall bar device). a,** Photocurrent map at $T$ = 14.2 K, where the photoresponse is dominated by potential fluctuations. **b,** Photocurrent map at $T$ = 4.2 K. **c,** The amplitude of the photocurrent decays with temperature, and it is observable for $T$ < 14 K. By contrast, the mean value of the conductance is $G \approx 1e^2/h$ independent of temperature. Scale bars, 1 μm. Experimental parameters are $E_{photon}$ = 1.5 eV.

**Supplementary Figure 6**

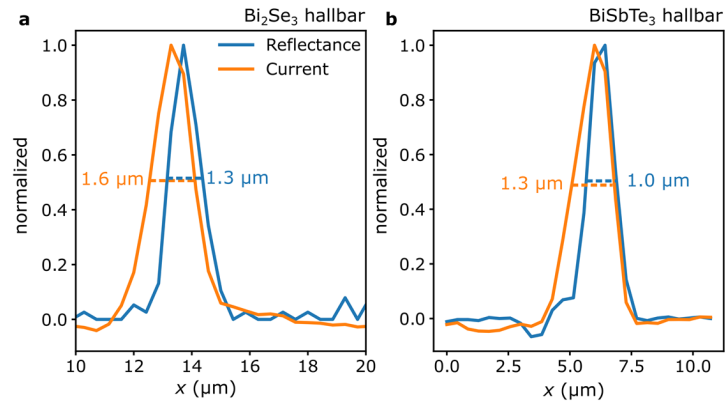

**Supplementary Figure 6. Spatial extent of the edge current.** Comparison of the positions of the edge current signal (orange line) and the physical edge of the circuit (blue line) for **a,** $Bi_2Se_3$ Hall bar and **b,** $BiSbTe_3$ Hall bar device. The position of the physical edge was determined from the gradient of the concurrently recorded reflectance image. The full width half maximum value of the reflectance is given by the laser spot size. Comparing the lateral extent of the current signal (orange line) and the reflectance signal (blue line) yields an estimate of the lateral size of the edge current on the order of 300 nm. Experimental parameters are $E_{photon}$ = 1.5 eV, $T$ = 4.2 K.
23